\documentclass{aa}
\usepackage{graphicx}
\begin{document}

\title{Planets orbiting Quark Nova compact remnants}

\author{P. Ker\"anen\inst{1}
\and R. Ouyed\inst{2,1}
}

\institute{
Nordic Institute for Theoretical Physics, Blegdamsvej 17,
DK-2100 Copenhagen \O , Denmark
\and 
Department of Physics and Astronomy, University of Calgary, 
2500 University Drive NW, Calgary, Alberta, T2N 1N4 Canada
}

\offprints{keranen@nordita.dk}

\date{Received/Accepted}

\abstract{
We explore planet formation in the Quark Nova scenario. 
If a millisecond pulsar explodes as a Quark Nova, a protoplanetary
disk can be formed out of the metal rich fall-back material. The propeller mechanism 
transfers angular momentum from the born quark star to the disk that will
go through viscous evolution with later plausible grain condensation
and planet formation. As a result, earth-size planets on circular orbits 
may form within short radii from the central quark star. The planets
in the PSR1257+12 system can be explained by our model if the 
Quark Nova compact remnant is born with a period of $\sim 0.5$ ms following the 
explosion. We suggest that a good portion of the Quark Nova remnants may 
harbour planetary systems. 
}
\maketitle

\section{Introduction}

The first planetary system beyond our own was found 
orbiting the millisecond pulsar PSR~1257+12 (Wolszczan \& Frail~1992; Wolszczan 1994). 
There are observations of three planets, $M_1=0.015 M_\oplus/\sin i_1$,
$M_2=3.4 M_\oplus/\sin i_1$ and $M_3=2.8M_\oplus/\sin i_3$, with the
semimajor axes $r_1=0.19$~AU, $r_2=0.36$~AU and $r_3=0.47$~AU, respectively. 
Here $i_i$ is the inclination of the orbit with respect to the observer.
The planets have very small eccentricities, 
i.e. their orbits are practically circular. 
The nature and origin of these planets is still unknown and debated
(see e.g. Lin et al. 1991; Phinney \& Hansen 1993; 
Podsiadlowski 1993; Miller \& Hamilton 2001). 
Models can be divided roughly into two classes depending on whether 
the planets are formed before or after the supernova (SN) explosion. 
In the pre-SN scenarios the planets form in a planetary disk before 
their host star exploded as a supernova. In the PSR~1257+12 system
the three innermost planets would have been inside the envelope of the
progenitor star before the explosion, evaporating or spiraling to the center
of the star. Planets surviving the supernova explosion could have been
far away from the star, and they could have migrated to their present
distances. It is argued that this scenario is not very probable (Miller \&
Hamilton 2001). It has also been argued that the explosion would unbind 
planets initially present, and any remaining planets
would have high eccentricities (Podsiadlowski 1993).

More attention has been focused on post-SN scenarios.
Summarizing the work of Miller \& Hamilton (2001),
it was found that if the planets form before or during the possible spin-up
period of the pulsar (following accretion from a companion star
which may also provide the disk material) 
they will be evaporated by the accretion luminosity 
unless the disk is thick enough.
If the planets are formed after the spin-up they must
be created from some remnant disk, but then the particle luminosity
from the neutron star (NS) is sufficient to disperse a tenuous disk material
faster than it can be supplied.
Added to that, the disk must be optically thick to the
high energy particles of the pulsar wind as to allow for planet formation. 
That is, the disk must be massive
enough $\sim 10^{28}$~g at a radius of  
$0.19$~AU 
(the radius of the innermost planet in the PSR~1257+12 system; 
see Miller \& Hamilton 2001 for more details).
The SN recoil scenario (Phinney \& Hansen 1993) where
the NS receives a kick in the direction of the companion as to capture
enough matter from the companion is favored.
This scenario also implies that many isolated NSs (maybe
up to around 10~\%) may have planets in orbit around them and 
must thus await further observational evidence.

In general, it seems that very special kind of circumstances
are required to produce planets around a pulsar.
Any model must demonstrate that the disk can expand
to the appropriate radius with suitable conditions for planet
formation on reasonable time scales (Bodenheimer 1993). Furthermore,
the mechanism cannot be too efficient, or otherwise more planetary 
systems would have been observed around the other nine isolated 
millisecond pulsars in the Galactic disk.
In this paper, we offer an alternative model related to the Quark Nova (QN)
phenomenon (Ouyed et al. 2002; Ouyed et al. 2003a).
In the QN explosion a protoplanetary disk forms from the fall-back matter.
The disk matter gains angular momentum from the central quark
star (QS) via the propeller mechanism. The effect of viscosity
on the disk evolution later allow for planet formation.
We describe these features and show how the points
mentioned earlier are dealt with. 

This letter is presented as follows: In Sect. 2. we review
the concept of QN. Sect. 3. deals with the angular momentum transport
and the disk formation. The evolution of the disk is described
in Sect. 4 where planetary formation is briefly discussed. We conclude
by applying our model to the PSR1257+12 system in Sect. 5.

\section{Quark Nova}

In the QN scenario (see
Ouyed et al. 2002 and Ouyed et al. 2003a) 
the quark core of a neutron star (NS) shrinks to the corresponding
stable compact/quark object before the contamination of the entire star.
By contracting, and physically separating from the overlaying
material (hadronic envelope which is mostly made of crust material), the 
core drives the collapse (free-fall) of the  left-out  matter leading to
both gravitational energy and phase transition energy release as high as
$10^{53}$~ergs. The QN-ejecta consist mainly of the NS metal-rich material 
outer layers. The accretion rate of the corresponding 
fall-back material is (see Ouyed et al. 2003a) 
\begin{eqnarray}
\dot{m}  
&\simeq &
10^{28}\, {\rm g/s}\, \big(\frac{\rho_{\rm ff}}{10^{6}\ {\rm g/cm^3} }\big)  
\big(\frac{R}{10 \, {\rm km}}\big)^{3/2} 
\big(\frac{M}{1.5\, M_{\odot}}\big)^{1/2}\, ,
\end{eqnarray}
where $\rho_{\rm ff}$ is the density of fall-back matter ($10^6$~g/cm$^3$, 
representing the neutron star crust material), $R$ is the radius of the QS
and $M$ its mass. 

%
%
%
%
The rotational period of the newly formed QS can
easily spin up to millisecond periods (here we adopt 
a period of $P=2$ ms)
and acquire magnetic field of the order of $B\simeq 10^{13}-10^{14}$~G.
The newly born QS is defined by three characteristic  radii:
the Keplerian ``co-rotation radius''
\begin{equation}
R_{\rm c}= 31\, {\rm km}\, \big(\frac{M}{1.5\, M_{\odot}}\big)^{1/3}
\big(\frac{P}{2\, {\rm ms}}\big)^{2/3}\, ,
\label{eq2}
\end{equation}
the magnetospheric radius at which the ram pressure
of the in-falling matter balances the magnetic pressure
\begin{eqnarray}
R_{\rm m}&=& \big( \frac{B^2R^6}{2\dot{m}\sqrt{2GM}}
\big)^{2/7}=78.4\, {\rm km}\, 
\big(\frac{B}{10^{14}\, {\rm G}}\big)^{4/7}
\times \nonumber \\ &\times &\big(\frac{10^{28}\,
{\rm g/s}}{\dot{m}}\big)^{2/7} \big(\frac{R}{10\, {\rm
km}}\big)^{12/7}\big(\frac{1.5\, M_{\odot}}{M}\big)^{1/7}\, ,
\label{eq3}
\end{eqnarray}
(see, e.g. Frank et al. 1992),
and the light cylinder
\begin{equation}
R_{\rm lc} = \frac{c}{\Omega} = 100\, {\rm km}\, \big(\frac{P}{2\, {\rm ms}}\big)\, .
\label{eq4}
\end{equation}
Given our fiducial values, the QS is born in the propeller regime, 
i.e. $R_{\rm c} < R_{\rm m} < R_{\rm lc}$ (Schwartzman 1970; 
Illarionov \& Sunyaev 1975), where the infalling
material may be accelerated in a wind that carries away
angular momentum from the magnetosphere and hence from the QS
itself. It is plausible that a reasonable fraction of QNe remnants
undergo a propeller phase after the explosion.

%
%
%
The angular momentum propelled away per unit time is 
\begin{equation}
\dot{L}_{\rm prop.}=
1.9\times10^{45}\,{\rm erg}\big(\frac{\dot{m}}{10^{28}\ {\rm g/s}}\big) 
\big(\frac{R_{\rm m}}{78\, {\rm km}}\big)^2 
\big(\frac{2\, {\rm ms}}{P_{\rm i}}\big)\, .
\end{equation}
The total angular momentum released in approximately $100$ seconds 
(the propeller lifespan; see Ouyed et al. 2003a) is thus of the order of 
$1.9\times 10^{47}\,{\rm erg\cdot s}$ carried away by $10^{30}$~g of fall-back material.
We note that a QS period can be as small as $0.5$ ms with the limit
set by the Kepler frequency (e.g., Glendenning 1997).
In most cases the fastest spin period is smaller than 1 ms;
the limit on how fast a gravitationally bound star (like NS)
can rotate before being ripped off by centrifugal forces.
That is, the angular momentum extracted by the propelled matter 
can easily be as high as $\sim 10^{48}\,{\rm erg\cdot s}$ if $P_{\rm i}\sim 0.5$~ms. 
It is still only $5-10$\% of the total initial angular momentum of 
the QS. Almost maximal angular momentum transfer might be possible,
but that would require specific conditions: very large 
accretion rates (and therefore a large amount of fall-back matter) 
together with a short propeller time to be able to compete with energy 
losses by gravity waves.  

Using eq.(11) in Ouyed et al. (2003a), we estimate that in $100$ seconds, 
the QS with $P_{\rm i}=2$~ms would have spun down (due to the propeller) by
30\% to reach a $P_{\rm f}=2.6$~ms period. Given
the initial rotational energy of the QS  
\begin{eqnarray}
E_{\rm i,rot.}= 7.4\times 10^{51}\, {\rm erg} \, \big(\frac{M}{1.5\, M_\odot}\big) 
\big(\frac{R}{10\, {\rm km}}\big)^2 \big(\frac{2\, {\rm ms}}{P_{\rm i}}\big)^2 \, ,
\end{eqnarray}
this implies $2\times 10^{51}$ erg in rotational energy is 
lost to the propelled material.

\section{Torus formation}

In order to keep some material bound to the star
the propelled matter must first interact 
with the dense infalling matter, releasing energy and angular momentum to it. 
Indeed, the potential energy of
the propelled matter, 
$-GMm_{\rm prop.}/R \sim -10^{50}$~erg, is not small enough
in comparison to its rotational energy.
The most realistic view of the process is the following: the matter propelled
away will be subject to the infalling matter. From the numbers above
it is straightforward to see that propelled matter should
interact with 10-20 times more matter (from the surroundings) to remain bound.
This mixed matter (propelled and falling-back sharing the angular momentum)
through viscous effects evolves into a torus. 
The more accurate study of the angular momentum transfer between the
propelled matter and the surroundings 
would certainly require the use
of numerical simulations and is beyond the
scope of this paper.
Here, and in order to carry on with our investigation, we again suggest that the 
outcome of this process is a torus or a protoplanetary disk.

The evolution of the torus is
governed by the angular momentum transfer due to turbulent viscosity 
(Shakura and Sunyaev 1973) 
\begin{equation}
\nu=\alpha c_s H = \alpha \frac{c_s^2}{\Omega} = \alpha \Omega H^2 \, ,
\end{equation}
where the parameter $\alpha\simeq (H_t/H)(v_t/c_s)\leq 1$ includes all 
uncertainties of the turbulence. 
Here, $H_{\rm t}$ is the largest eddy size while $H$ is the vertical height of the 
disk. The radial diffusion timescale is $\tau_\nu(r)\simeq r^2/\nu$, and if the viscosity is 
expressed by the above $\alpha$ formula, 
the viscous evolution time of this torus is
$\tau_\nu(r)=\alpha^{-1}(r/H)^2\Omega^{-1}$ where $\Omega (r)$ is
the angular velocity at radius $r$. That is,
the torus will expand within some seconds to a distance of $10^3-10^4$~km
since $r/H\sim 1$ (the ratio of the disk radius and the scale height)
and assuming $\alpha\sim 0.01-1$.

In these early phases the accretion to the QS is large, 
so the viscous time scale for most of the torus to be accreted is at best some 
Keplerian orbital times.  Nevertheless, we expect the star to remain bare for 
two reasons: i) the propeller is still efficient at deflecting matter and 
ii) the star's surface temperature is very high ($> 10$ MeV). 
Therefore, it is natural to expect that any normal matter that
managed to evade the propeller is ejected away as a hot 
wind\footnote{The physics for calculations of gas outflow 
from the crust is similar for both NSs and QSs (Usov 1997).}.
The torus is also very dense and would likely survive 
the hot wind and any radiation field.  

\section{Disk Formation and Evolution}

The torus described above will further expand to form a disk.
The formation and evolution of such a disk is
also governed by the angular momentum transfer due to turbulent viscosity. 
While most of the matter is accreted to the QS, most of the
angular momentum is carried out with the part of
matter demigrated to some larger distances.

If one approximates $H\sim (0.01-0.1)r$ and $\alpha\sim 0.01$, the viscous 
timescale for the disk reaching $r=1$~AU is $\tau_\nu(r)\sim 10^3-10^5$~yrs.
The volume of a disk with $r\sim10^{13}$~cm and with a scale height 
$H\sim10^{12}$~cm is roughly $\pi r^2 H\sim10^{38}\, {\rm cm}^3$. With the total mass
of $10^{28}-10^{29}$~g, the average density is  $10^{-10}-10^{-9}$~g/cm$^3$, 
corresponding to an average proton density of $10^{14}-10^{15}$~cm$^{-3}$. The number density
of particles is expected to be lower since the disk is rich in heavy nuclei.

At later stages, the QS surface temperature would have
decreased allowing for crust formation (channelled along the field line).  
The kinetic energy of the subsequently accreted gas may be transformed into 
emission of the QS atmosphere in the
following way (e.g., Xu 2002): The magnetic field channels the gas motion along the 
field lines to the QS. In this case, the kinetic energy of
ions at the surface is about 100 MeV/nucleon, which is 5 times greater than the 
Coulomb barrier at the quark surface (Alcock et al. 1986). Accreted particles 
penetrate through the quark surface, and they are dissolved into quark matter. 
As a result, the quark core is heated at the magnetic poles. The process of heat 
transport through the core is very fast because of the very high heat conductivity 
of quark matter, and therefore the quark core is nearly isothermal. Then the energy 
that is released in the process of gas accretion is radiated from the
normal-matter atmosphere (the crust) more or less isotropically. However, this 
process occurs much later in the evolution, the disk then being geometrically 
thin enough to survive photodissociation as discussed below.

The luminosity can be estimated conservatively as follows: if the whole 
disk matter is accreted to the central object within reversed viscous 
timescale, the rate should be 
at most $\sim 10^{29}\, {\rm g}/10^5\, {\rm yr}\sim 3\times 10^{16}$~g/s, 
which is roughly 0.03 of the maximal rate corresponding to the Eddington luminosity. 
When accreted to the surface of the QS, the total released energy in radiation is 
$L=\eta \dot{m} c^2$, where $\eta$ is the accretion efficiency;
in our case conservatively $\eta\sim 0.15$ (it seems appropriate as a first approximation to take the accretion efficiency of QSs to be close to that of
black holes and of neutron stars). 
The QS luminosity, the disk geometry and viscous 
dissipation will define the temperature of the disk. The dissipation 
can be neglected considering the above mentioned accretion rates 
(see e.g. Ruden 1993).
The black body temperature of a perfectly absorbing plane with its surface 
inclined at an angle  $\tan\beta=H/2r=0.005$ to the 
radiation is $T=[\eta\dot{m}c^2\tan\beta/(\sigma 4\pi r^2)]^{1/4}\sim 1330$~K
at $r=0.2$~AU and less than $600$~K at $r=1$~AU (using $\eta=0.15$);
assuming $L_{\rm E}$ one obtains $3500$~K and $1600$~K, respectively.
Lower temperatures are possible since for
a rapidly rotating compact object where the accreted matter closely
corotates with the surface (the matter ``softly'' lands onto
the surface), as in our case, the efficiency
can be smaller ($\eta\sim 0.05$; Sibgatullin \& Sunyaev 2000).

Since the baryon density of the disk is 
$n\sim10^{14}-10^{15}$~cm$^{-3}$, the mean free path of photons 
is $\lambda=1/(n\sigma_{\rm T})\sim 10^9-10^{10}$~cm.
Therefore radiation does not penetrate the inner parts of the 
optically thick disk, and it can cool to the temperatures 
estimated above. Added to that, the high energy wind particles  
cannot penetrate it either, since the
Coulomb cross section is larger than the Thomson cross section. 
Therefore the disk can easily protect the forming dust grains 
(and later planetesimals) from the possible radiation of the QS. 
Iron and most of the other chemical components will condense into dust grains 
when the disk has cooled to temperatures below $1500$~K (see e.g. Lewis 1995). 
The condensation timescale for different dust grains (e.g. Sedlmayr 1993) 
are roughly some months up to some tens of years. The dense metal rich environment 
with viscous timescale that is longer than condensation timescale makes the 
QN disk suitable for dust grain formation. The growth of these grains and
later planetesimals (see e.g. Lissauer 1993; Ruden 1999) 
will lead to the plausible formation
of planetary bodies up to some earth masses, with small (up to some~AU)
circular orbits.

\section{The PSR1257+12 system}

Before applying our model to the case of PSR~1257+12, we note that
there has been some observational evidence of a planet or a companion 
orbiting the pulsar system PSR~1620-26, with a mass around 
$0.01\,M_\odot/\sin i$ and the semimajor axis of around $40$~AU (Joshi 
\& Rasio, 1997). This is however a triple system in a globular cluster
and the formation history might be very different from the one we are 
considering here (see e.g. models of Sigurdsson 1993; Ford et al. 2000).

The three confirmed planets of the PSR~1257+12 have a total 
mass of $\sim 4\times 10^{28}$~g.
Most of the disk material (more than 90\%) in our model 
would have fallen back to the QS leaving only 
$(0.01-0.1)\times 10^{30} = 10^{28}-10^{29}$~g orbiting the QS.
The total angular momentum of the PSR~1257+12 planets is 
$\sim 1.3\times 10^{48}\,{\rm erg\cdot s}$, which is an order of magnitude higher 
than our estimates (see Sect.~2.). This can be accounted for (without stretching 
the model) by considering a longer propeller life span ($\gg 100$~s) 
and/or a faster spinning QS ($P_{\rm i} \sim 0.5$~ms), which could then 
provide as much angular momentum as there is in the PSR~1257+12 
planets\footnote{However, if the speculated large outer planet does 
exist (Wolszczan~1996, Joshi \& Rasio~1997), the angular momentum will be 
far too little to explain its formation via the mechanism described here,
even if parameters are stretched to their limits.}.

The difficult challenge, however, is explaining PSR~1257+12
measured dipole magnetic field\footnote{PSR~1257+12 has a 
period of $P=6.219\times 10^{-3}$~s, and a period derivative 
of $\dot{P}=1.2\times 10^{-19}$, which implies a dipole 
magnetic field of $B=3\times 10^{19}(P\dot{P})^{1/2}\approx8.8\times 10^8$~G 
in the standard magnetic dipole radiation spin-down models.} which
is $B\approx 8.8\times 10^8$~G. With such a weak field the propeller would 
not have worked making it difficult to form a protoplanetary disk.
In our model, the importance of the propeller is two-fold: 
i) it supplies enough angular momentum for disk formation and 
ii) delays the formation of the crust thus drastically reducing 
accretion luminosity and stellar wind during the early stages of 
disk formation. One possibility is that the magnetic field has 
decayed. While this hypothesis would have difficulty if PSR~1257+12 is
a NS (see Camilo et al. 1994 for a discussion)
this is not necessarily the case if one assumes that PSR~1257+12 is  
a QS: an interesting feature of such an object is the plausible 
decay of the magnetic field due to the Meissner effect at its surface 
(Ouyed et al. 2003b), a notion which remains to be confirmed.  
We thus argue that PSR~1257+12 might be a QN compact remnant 
(born as a millisecond QS) which experienced such a phenomenon.
Finally, comparing the QN rate of 1 per million years per galaxy 
to that of the SN rate of $2\times 10^{4}$ per million years per 
galaxy, on average 1 in $2\times 10^{4}$ compact 
objects are QN remnants. We suggest that a good portion of this
QS population may harbour a planetary system as described in this work.

\begin{acknowledgements}
We are grateful to A.~C.~Andersen, G.~Bj\"ornsson, A.~Brandenburg, 
E.~H.~Gudmundsson and J.~Poutanen for helpful discussions.
We thank H.~Dahle for bringing the PSR~1257+12 system to our
attention, and the Science Institute of the University of Iceland for hospitality. 
\end{acknowledgements}

\end{document}